# Automated technique to reduce positive and negative false from attacks collected through the deployment of distributed honeypot network

[1]Abdeljalil AGNAOU, [1]Anas ABOU EL KALAM, [1]Abdellah AIT OUAHMAN, [2]Mina DE MONTFORT

[1]OSCARS Laboratory - National School of Applied Sciences Marrakech, Morocco
[2]ARTIMIA, Paris FRANCE
agnaou.abdeljalil@gmail.com, elkalam@hotmail.fr, aitouahman@yahoo.com, mdemontfort@artimia.fr

*Abstract*- **Current tools and systems of detecting vulnerabilities simply alert the administrator of attempted attacks against his network or system. However, generally, the huge number of alerts to analyze and the amount time required to update security rules after analyzing alerts provides time and opportunity for the attacker to inflict damages. Moreover, most of these tools generate positive and negative falses, which may be important to the attacked network. Otherwise, many solutions exist such as IPS, but it shows a great defect due, fundamentally, to false positives. Indeed, attackers often make IPS block a legitimate traffic when they detect its presence in the attacked network.**

**In this paper we describe an automated algorithm that gives the ability to detect attacks before they occurrence, then reduce positive and negative falses rates. Moreover, we use a set of data related to malicious traffic captured using a network of honeypots to recognize potential threats sources.**

I. Introduction:

During the last years, many proposals for the use of honeypots were held. Some of them were deployed to waste the time of the attackers, others to reduce the activities related to spam, or to just deceive attackers, and some others to analyze the intrusion of the hackers.

The applications of the honeypots are in hence completely diverse, the most important approaches are first presented and studied with details in this paper.

In the context of our own experiences, we have created a distributed Honeypots-based platform as well as some tools for intrusion detection and vulnerabilities analysis, designed to monitor malicious traffic and prevent attacks in a long-term perspective. Deployed across different partners in different organizations, tour platform is firstly used to obtain statistics on attacks against information systems of the collaborators of our project, and then, to correlate results in order to reduce false positives (FP an alert for an event that is not a threat) and false negatives (FN an alert for an event which has not been detected but is a real threat) Note that FP and FN is a bug challenge for existing approaches and tools.

Actually, combining several tools and techniques could allow remedying this problem in some cases. For example, if vulnerability is detected by a number of tools, the probability of occurrence may be high.

Subsequently, the first aim of this paper is to provide an intelligent algorithm that can aggregate multiple techniques and / or vulnerability management tools, in order to have a more effective and evidential results, leading to better detection rates and reducing false positives and false negative.

To achieve these ends, we propose through to use artificial intelligence as well as distributed network of honeypots-based techniques. We also propose aggregation algorithms for the evaluation, testing and benchmarking. Indeed, the honeypot network could help to have statistics on attacks, their origins, behaviour of attackers, etc. will artificial intelligence could help to provide mathematical foundation to define, specify and evaluate models that we will set throughout this article.

The reminder of this paper is organized as follows: Section 2 discusses related works. Then, the Section 3 gives an overview about Data Collection Environment used as a basis to implement our intelligent algorithm, in the section 4 we present our algorithm with all its components, Section 5 details some results for the evaluation of our algorithm, Finally section 6 conclude this paper and presents our project prospects.

I. Related works:

With the continuous development of honeypots in recent years, and deployment of new distributed honeypot-based platforms around the world, the amount of information collected becomes extremely large. Consequently, in several honeypot platforms, security administrators become overwhelmed with the large amount of data to analyse. Therefore, many researchers have recently started offering automated analysis solutions based on intelligent analysis methods such as machine learning, data mining and statistical tools.

Pouget and Dacier [1] proposed a simple approach of clustering to analyze data collected from the project of distributed honeypots "Leurre.com". Their goal was to characterize the root causes of attacks on the honeypots. The purpose of this algorithm is to collect all the attacks with some common characteristics (duration of attack, targeted ports, the number of packets sent, etc. ...) based on generalization techniques and extracting association rules. The resulting clusters are more refined using "the Levenshtein distance". The ultimate goal of their approach is to group into clusters all sources of attacks sharing similar activity footprints, or attack tools.

Alata et al. [2] proposed simple models describing the time evolution of the number of attacks observed on different platforms of honeypots [Refs …]. In addition, they studied the potential correlation of attack processes observed on these different platforms, taking into account the geographical location of the attackers and the relative contribution of each platform in the overall attack scenario. The correlation analysis is based on a linear regression model.

In [3], Dacier and Thonnard have proposed a Framework for the discovery of attack patterns in the honeynet data; the aim of this approach is to find, in a set of attack data, network traces groups sharing various similar models. In this work, the authors applied a method of classification based on the graphs to analyse one specific aspect (the set time of the attacks) of the honeynet data. The results of the groupings applied to the analysis of time series have allowed identifying the activities of several worms and botnets in traffic collected by honeypots.

To facilitate analysis of data collected by honeypots and identification of attacks, members of the "Honeynet Project"[4] suggested the use of a tool called Picviz [5]. Picviz is a parallel coordinate plotter which allows acquisition of data from various inputs (tcpdump, syslog, iptables logs, apache logs, etc ...) and then ,visualize and discover interesting results. This tool enables the representation of complex multidimensional events in a two-dimensional interface only.

Seifert et al. [6] proposed static heuristics method to classify malicious web pages. To implement this method, they used machine learning to build a decision tree to classify web pages as normal or malicious. Subsequently, malicious Web pages are sent to a high-interaction honeypot for a second inspection. The goal of this method is to reduce the number of web pages, which will be inspected by the honeypot.

Table 1 summarizes the basis (overview, data source environment, analysis techniques) of some of existing works.

TABLE 1
BASIS OVERVIEW OF SOME EXISTING WORKS

| Reference | General idea | Data Source Environment | Analysis techniques used |
|---|---|---|---|
| Pouget and Dacier et al. | Caracterize root causes targeting honeypots | Leurre.com | Association rules Clustering |
| Alata E and al. | Provide models based on observed attacks | Leurre.com | Correlation, Linear Regression |
| Thonnard and Dacier | Find attack groups sharing diverss kinds of similarities to explore the underlying causes | Leurre.com | Clustering based on graphs |
| Honeynet Project | Visualization of data received by honeypots and simplify its presentation | Honeynet | Parallel coordinates |
| Seifert and al. | Classify pages as normal or malicious | Honeypot client for the web | Heuristic static, decision tree |

In this paper, we propose an approach that shares the same objective with the other works that we have described previously, with more facilities analysis of data collected from the distributed network of honeypots. Moreover, our approach offers more new features in comparison to other works. In fact, we offer an intelligent algorithm, dedicated to the treatment of security events, and enable the reduction of false positives and false negatives, lacking in the literature. The proposed module combines multiple vulnerability management methods, and is boosted by artificial neural networks; which simplifies the supervisory task for the security administrator. Our algorithm is evaluated through several experiments to show its performances.

II. Data Collection Environment: Moroccan Honeypot Project

This section describes data collection environment used for the scope of our project: the Moroccan Honeypot Project (MHP) how data is collected as well as its internal structure.

Basically, the "Moroccan Honeypot Project" aims to provide to Moroccan scientific community of researchers a platform easy to deploy, allowing them collect and analyse attacks and their distribution as well as the hacker's profiles, and giving administrators the required time to better protect their networks before being victim of a potential attack. Such procedure would decrease remarkably risks related to sophisticated attacks, because it will be analysed by humans who will decide if the reported activity really represents a threat or not. It is to highlight that MHP project is the first of its kind in Morocco. It represents an interesting way to quantify precisely the amount of malicious traffic occuring against information systems in Morocco, which is classified, according to a report in 2015, as the $3^{rd}$ targeted county in Africa. Such projects becomes a necessity because of its advantages regarding security issues and cybercrime evaluation.

The MHP manages multiple types of honeypots and information sources. It manipulates many types of High interaction honeypots such as Dionaea [7], kippo [8] and P0f [9], as well as the famous IDS Snort [10], consisting of physical machines running real operating systems (UBUNTU 14.04). The MHP honeynet supports multiple subnets consisting of IP addresses contributed by different organizations participating in the research. The main server collects multiple sources of information from different distributed devices (e.g., Snort events from Snort Sensors, and malware from Dionaea, OS fingerprints from P0f, passwords from kippo and more), analyses the data, and presents it to users in an efficient and actionable manner.

MHP is based on the Modern Honey Network [11], which is an open source Framework for Management of distributed honeypots developed by ThreatStream. It provides deployment and events aggregation capabilities for several of the current open source honeypots available; it supports external and internal honeypot deployments at a large and distributed scale. It is used as a basis for the MHP project in addition to several open source tools such as the famous SIEM "Splunk".

There are many factors behind choosing MHN as a basis of the MHP Project. Firstly, Honeypots are allocated from MHN central server in the distributed sensors, which sent in turn the collected data to a central server acting as processing engine and database server. Secondly, the damage impact is lower and only limited in honeypots sensors. This characteristic is important, as it's possible to cause loss or destruction of information or system's integrity, so this issue is highly considerable in MHN architecture.

Basically, from MHP, there are two key functionalities required to mention: event aggregation capabilities and honeypot installation and configuration scripts. The installation and configurations scripts are written in bash language and include support for several types of honeypots, such as: Dionaea, Kippo, Wordpot, Snort and many others.. The data aggregation involves the centralization of the captured malicious traffic from the network of sensors. Each honeypot is specialized in a kind of data, for example Kippo presents most used users and passwords, Dionaea aims to trap malwares exploiting vulnerabilities exposed by services offered over a network, and ultimately obtain a copy of the malware, etc. These data includes also source IP, attacked port, associated protocol and geo-location with a real time map.

To understand how it works, the MHP's architecture is structured in six layers, as shown in figure 1.

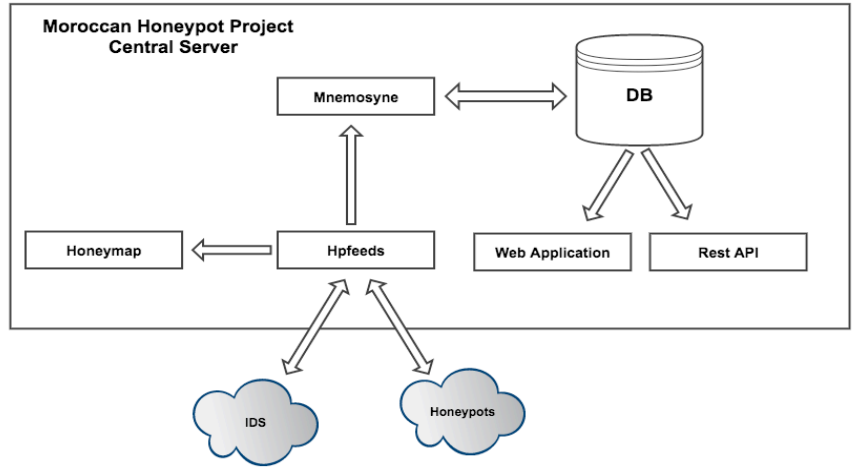

Figure 1. The MHP's Architecture.

The first layer is the Honeypot sensors. When it's compromised, data related to attacks is sent to the MHP central server, this operation is made by Hpfeeds [12].

Hpfeeds, The second layer, is a protocol ensuring the creation of a secure channel for transmitted data. Hpfeeds is actually a lightweight authenticated publish-subscribe protocol largely used with honeypot technology.

Although the data is received into the Platform, we're in the phase of processing; Mnemosyne [13] is the component that deals with it. It has three main objectives: (1) providing immutable persistence for hpfeeds, (2) offering the normalization of data, and finally (3) exposing the normalized data through a Rest API.

Finally, the last layer consists in a web application, serves as user interface to consult the data from database. It is implemented with Flask, which is a micro-framework that combines a frontend implemented with HTML templates and a backend in Python.

Another tool used is honeymap, which allow us to have real time map characterising geo-location of attacks occurring against our sensors.

Captured data details and their use in analyzing security issues are discussed in the next sections.

### III. Data analysis Framework

#### A. Overview and main architecture

In this section, we will present the main contribution of this paper, which is modeling an intelligent and automatic technique to reduce FP and FN. Our algorithm is be based on several methods of detecting vulnerabilities, and boosted by the neural networks, with the main objective to simplify the task of security administrator. This algorithm will be implemented in Rest API layer on the MHP server.

The intelligent algorithm based on the fact of using several sensors installed within the same site and being able to find the same attack signature. The architecture of a site is shown in the Figure 2:

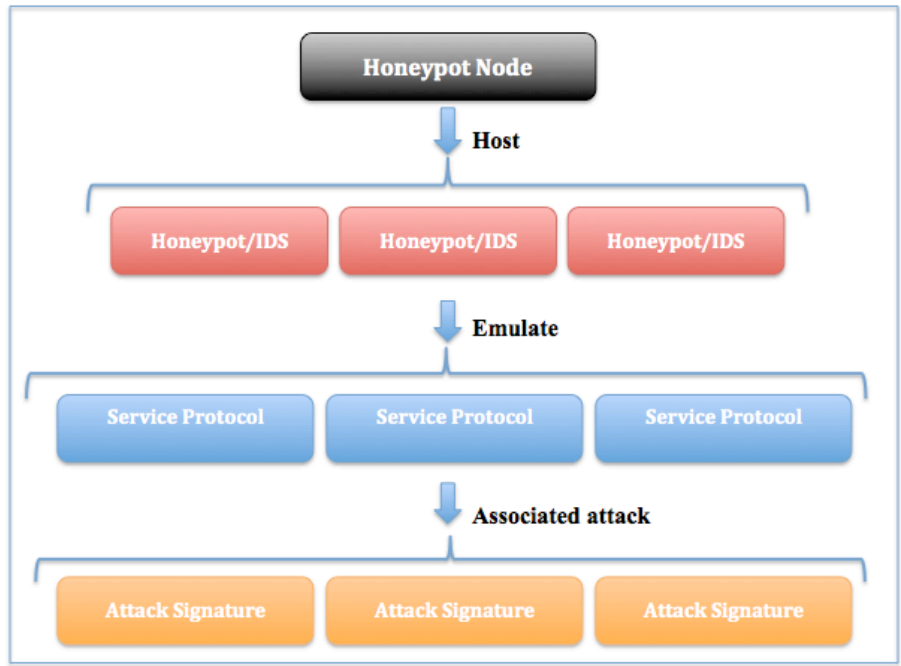

Figure 2. Detailed diagram of a member site.

Basically, as none of the deployed sensors (honeypot or IDS) is perfect (e.g., each one of them commits false negative and generates false positive), we use a probabilistic model for calculating credibility parameters of each sensor; and based on these parameters, the artificial neural network [14] decide whether an alert is associated with a real attack or not.

As we use the Moroccan Honeypot Project as Data collection environment, our Framework will be implemented within the architecture of the MHP; Figure 3 shows the integration of our system with its components:

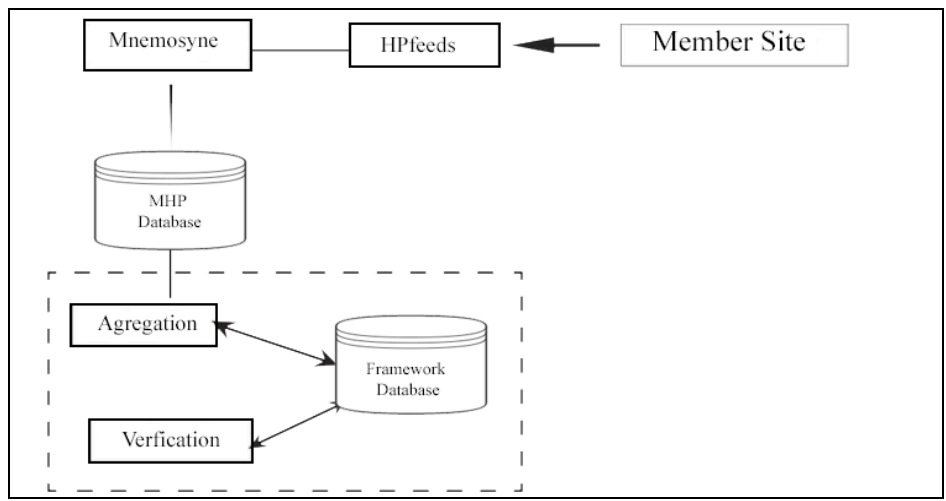

Figure 3. Integration of the intelligent framework into MHP Architecture.

Firstly, once data is collected from a member site through Hpfeed's channels, Mnemosyne in engaged to normalize and store it into MHP database. The next step is then proceeding to the analysis of this data (which is achieved by our framework).

The last three components makes our framework: (1) the aggregation component allows to gather the sessions of attacks following certain criteria into sets called Meta-alerts, (2) the framework DB component is a MySQL database that stores the data necessary for the proper functioning of the two components, aggregation and verification, and also to accommodate the meta-alerts verified by the (3) verification component.

These components are described more in details in the rest of this Section.

### B. The Aggregation Component

The aggregation component allows us to gather the alerts associated to the same attack session, that is to say the same timestamp, the same socket and the same attack signature.

Aggregation provides as output a meta-alert that contains both the information about attack session concerned, the list of sensors that have generated alerts, and also the list of sensors in the same member site, which may help to detect the attack signature concerned and have not generated alerts.

Basically, a Meta-alert associated with a signature can be modelled mathematically by a combination $(a_1, ..., a_i, ..., a_N)$ Written under the base *B* as:

- if $a_i = 1$: the sensor with the index 'i' has issued an alert.

- if $a_i = 0$: the sensor with the index 'i' has not issued anything.

- N: number of sensor that can detect the signature.

- B: (sensor 1, ..., sensor $i$, ..., sensor $N$).

The aggregation component periodically collects all the standardized alerts with a timestamp greater than the date of the last collection from layer 4 of the MHP, which is the database. Then (after a meta-alert is closed), the meta-alert is forwarded to the audit component.

```python
def merge(self,session_object) :
    self.HPalert.append(session_object.honeypot)
    c=0
    for i in range(len(self.HPnoalert)) :
        if self.HPnoalert[i] == session_object.honeypot:
            c=i
            break
    ll = self.HPnoalert.pop(c)
    if len(self.HPnoalert) == 0 :
        self.open=0
        self.Ptrue = 1  #meta-alert is defined as a correct alert
        self.tag = 1
    self.sessions.append(session_object)
```

Figure 4. Method of the meta-alert class responsible to aggregate sessions.

### C. The Verification component

The verification component is the intelligent component of our framework; it is thus responsible for deciding whether a meta-alert has a "Real threat" or a "False threat".

The functioning of this component is divided in two phases, the first one called the *learning phase*, in which we calculate the parameters allowing the verification. The second one called real time recognition, in which the verification component uses the parameters calculated during the learning phase to classify the generated meta-alerts by the aggregation component.

As mentioned earlier, a meta-alert associated with an attack signature is translated in a combination $(a_1, ..., a_i, ..., a_N)$. Based on the *N* binary values of this combination the following two probabilities are calculated which represent the two significant values provided to the artificial neural network which handles the task of decision:

- **Ptrue**: is the probability of the event "the meta-alerts present a real threat." That is to say that all sensors that have generated alerts have "true positive" and that all sensors that have nothing generated have "False Negative".

- **Pfalse**: is the probability of the event "the meta-alerts present a false threat." That is to say that all sensors that have generated alerts have "false positive" and that all sensors that have nothing generated have "negative true".

Knowing that the type of traces that presents a sensor is independent of the type of traces that do other sensors, the calculation of these two probability formulas are (1) and (2):

- Ptrue = $\prod_{i=1}^{N}(a_i \cdot P(X_i = TP) + (1 - a_i) \cdot P(X_i = FN))$     (1)

- Pfalse = $\prod_{i=1}^{N}(a_i \cdot P(X_i = FP) + (1 - a_i) \cdot P(X_i = TN))$     (2)

Bayes' theorem is defined as following in (3):

$$P(A_k/B) = \frac{P(B/A_k) \cdot P(A_k)}{\sum_{1}^{n} P(B/A_i) \cdot P(A_i)}, k=1...n \quad (3)$$

TABLE 2
DESCRIPTION OF THE DIFFERENT NOTATIONS USED.

| Notation | Description |
| --- | --- |
| M: {0,1} | 1: trace is malicious, 0: if not |
| A: {0,1} | 1: Alert is generated, 0: if not |
| p: {http, ssh...} | Used protocol to classify traces |
| Type: {TP, FP, TN, FN} | Type of the trace |
| $X_i$ | Type of trace that represents the $i^{th}$ sensor |
| N | Number of sensors associated to the meta-alert |
| $R_{p,Type}^{S}$ | Generation rate of the type of traces under the protocol p and the signature S |
| $R_{p,Type}$ | Generation rate of the type of traces under the protocol p |

For the $i^{th}$ sensor of the meta-alert associated with the signature 'S' under the protocol 'P', the following probabilities using Bayes theorem are calculated:

$$P(X_i = TP) = P(M=1/A=1) = \frac{P(A=1/M=1) \cdot P(M=1)}{P(A=1/M=1) \cdot P(M=1) + P(A=1/M=0) \cdot P(M=0)} \quad (4)$$

$P(X_i = FP) = P(M=0/A=1) = 1 - P(M=1/A=1) = 1 - P(X_i = TP)$     (5)

$$P(X_i=TN) = P(M=0/A=0) = \frac{P(A=0/M=1).P(M=1)}{P(A=0/M=0).P(M=0) + P(A=0/M=1).P(M=1)} \quad (6)$$

$$P(X_i=FN) = P(M=1/A=0) = 1 - P(X_i=TN) \quad (7)$$

(4) (5) (6) (7) represents respectively the probabilities that the sensor 'i' present in the present time TP, FP, TN, FN. Similar to P (A = 1 / M = 1) P (A = 1 / M = 0), P (A = 0 / M = 0), P (A = 0 / M = 1), they represent respectively the probabilities that the sensor 'i' present in the past time TP, FP, TN, FN. These conditional probabilities are defined as following:

$$P(A = 1/M = 1) = \begin{cases} R_{p,TP}, & si\ R^S_{p,TP} = 0 \\ R^S_{p,TP}, & sinon \end{cases} \quad (8)$$

$$P(A = 1/M = 0) = \begin{cases} R_{p,FP}, & si\ R^S_{p,FP} = 0 \\ R^S_{p,FP}, & sinon \end{cases} \quad (9)$$

$$P(A = 0/M = 1) = \begin{cases} R_{p,FN}, & si\ R^S_{p,FN} = 0 \\ R^S_{p,FN}, & sinon \end{cases} \quad (10)$$

$$P(A = 0/M = 0) = R_{p,TN} \quad (11)$$

Finally, P (M = 1) represents probability of malicious trace for signature 'S' under the protocol 'P'. All of these five probabilities will be calculated during the learning phase.

*1) The Training Phase*

Before beginning each training phase, we first must have:

- A training traffic composed of benevolent traces and other malicious,
- The number of benevolent traces under each protocol,
- The number of malicious traces for each of the attack signatures.

The learning phase begins with the exposure of the training traffic to the sensors included in the test. Once alerts generated by the sensors are collected by "HPfeed" [12], normalized and indexed by "Mnemosyne" [13] and stored in MHP DB, they should be checked by the security administrator and then copied into the 'Framework DB'. Based on this verification, we calculate for each sensor the $R_{p,TP}$, $R_{p,FP}$, $R_{p,TN}$ and $R_{p,FN}$ related to each Protocol 'p', the percentage $P(M = 1)$ of malicious traces under 'p', and $R^S_{p,TN}$, $R^S_{p,FP}$, $R^S_{p,FN}$ related to each signature 'S' under the protocol 'p'. These rates are calculated by the following formula and stored in the table "rate" of the Framework DB database (see next section):

$$R^S_{p,TP} = \frac{\text{Number of alerts having the signature 'S' verified as real threat}}{\text{Total number of alerts received having signature 'S'}} \quad (12)$$

$$R^S_{p,FP} = 1 - R^S_{p,TP} \quad (13)$$

$$R^S_{p,FN} = 1 - \frac{\textit{Nombre d'alertes de signature 'S' vérifiées comme vrai menace}}{\textit{Nombre de traces malveillantes de signature 'S'}} \quad (14)$$

$$R_{p,TP} = \frac{\text{Number of alerts under the Protocol 'p' verified as a real threat}}{\text{Total number of alerts received under the protocol 'p'}} \quad (15)$$

$$R_{p,FP} = 1 - R_{p,TP} \quad (16)$$

$$R_{p,FN} = 1 - \frac{\text{Number of alerts under the Protocol 'p' verified as a real threat}}{\text{Number of malicious traces under the protocol 'p'}} \quad (17)$$

$$R_{p,TN} = 1 - \frac{\text{Number of alerts under the protocol 'p' verified as false threat}}{\text{Number of benevolent traces under the protocol 'p'}} \quad (18)$$

$$P_p(M = 1) = \frac{\text{Number of malicious traces under the protocol 'p'}}{\text{Total number of traces under the protocol 'p'}} \quad (19)$$

The next step is to merge received alerts to create meta-alerts, which will be classified following the attack signatures, to train the multilayer Perceptron (see next) according to different signatures. It means that we want the multilayer Perceptron to have a behaviour that depends on the attack signature.

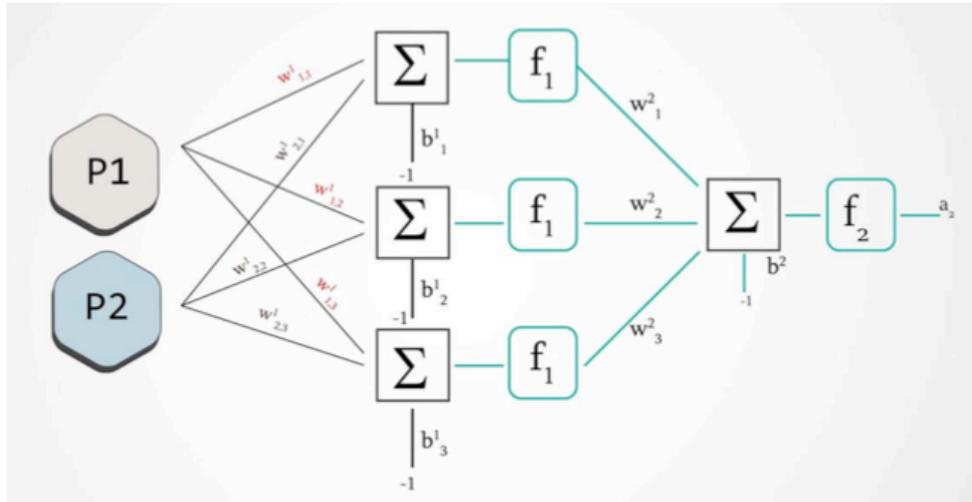

*Figure 5: Multilayer perceptron implemented*

To make the task of supervised learning, we put into the input of the neural network the set of pairs {(P1, D1), (P2, d2)... (PQ, dQ)} (Training Pattern), such as:

- Pi: Is the vector [Ptrue, Pfalse] of the i[th] meta-alerts.

- Di: Is the value we want to associate to Pi, it is equal to 1 (for meta-alerts pre-verified as real threat) or 0 (for meta-alerts pre-verified as false threat).

The training algorithm of the framework is presented by the following pseudo code:

1. Verify and store sessions.

2. Calculate for each sensor rate by signature and by protocol

3. Merge sessions into meta-alerts

4. Calculate the two meaning probabilities for each meta-alert.

5. Execute the training algorithm for the multilayer perceptron.

*2)* *The Real-time Phase*

After calculating weight matrix associated with each signature in the learning phase, we move now to the real-time operation of our network, in which the network decides the positivity of a meta-alert based on the pre-calculated weights. The verification function first calculates the probabilities of each two significant meta-alert for which there are sensors that have not generate alerts, and then passed the two significant values (Ptrue and Pfalse) in the neural network.

```python
def verify(self,cnx) :
    curser=cnx.cursor()
    rates = [] #Rates allowing to calculate the significant probabilities
    for i in range(0, len(self.HPalert)) :
        rates.append(rate(1, self.sig_id, self.sessions[i].proto_id,curser))
    for i in self.HPnoalert :
        rqt10 = "select protocol.id from protocol, honeypot, proto_sig where proto_sig.sig_id = '"+str(self.sig_id)+"' and proto_sig.proto_id = protocol.id and protocol.hp_id = honeypot.id and honeypot.name = '"+str(i)+"'"
        curser.execute(rqt10)
        proto_id = curseur.fetchall()[0][0]
        rates.append(rate(0, self.sig_id, proto_id, curser))
    self.Ptrue =1
    self.Pfalse =1
    x = 0.0
    for r in rates :
        if r.v == 1 :
            x = r.rtp * r.pn / (r.rtp * r.pn + r.rfp * (1 -r.pn))
        else :
            x = r.rfn * r.pn / (r.rfn * r.pn + r.rtn * (1 - r.pn))
        self.Ptrue *= x
        self.Pfalse *= (1 - x)
    if self.test == 0 :
        n = NN(2, 3, 1)
        rqt11 = "select w1, wo from weight where weight.sig_id = '"+self.sig_id+"'"
        curser.execute(rqt11)
        o=curser.fetchall()
        W = []
        for d in o:
            W.append(d[0])
        n.wi = stlist(W[0],3,3,1)
        n.wo = stlist(W[1],3,1,1)
        if n.update([self.Ptrue, self.Pfalse]) > 0.5 :
            self.tag = 1
        else :
            self.tag = 0
```

Figure 6. Verification function of meta-alerts.

If the output of the neural network is strictly greater than the value '0.5', the audit function gives a label of 'real threat' to the meta-alert, otherwise the function tag meta-alert as a 'false threat'. Therefore, the accuracy of the decisions depends on the accuracy of the training data. The Framework database where all the results are stored, is presented in the figure below:

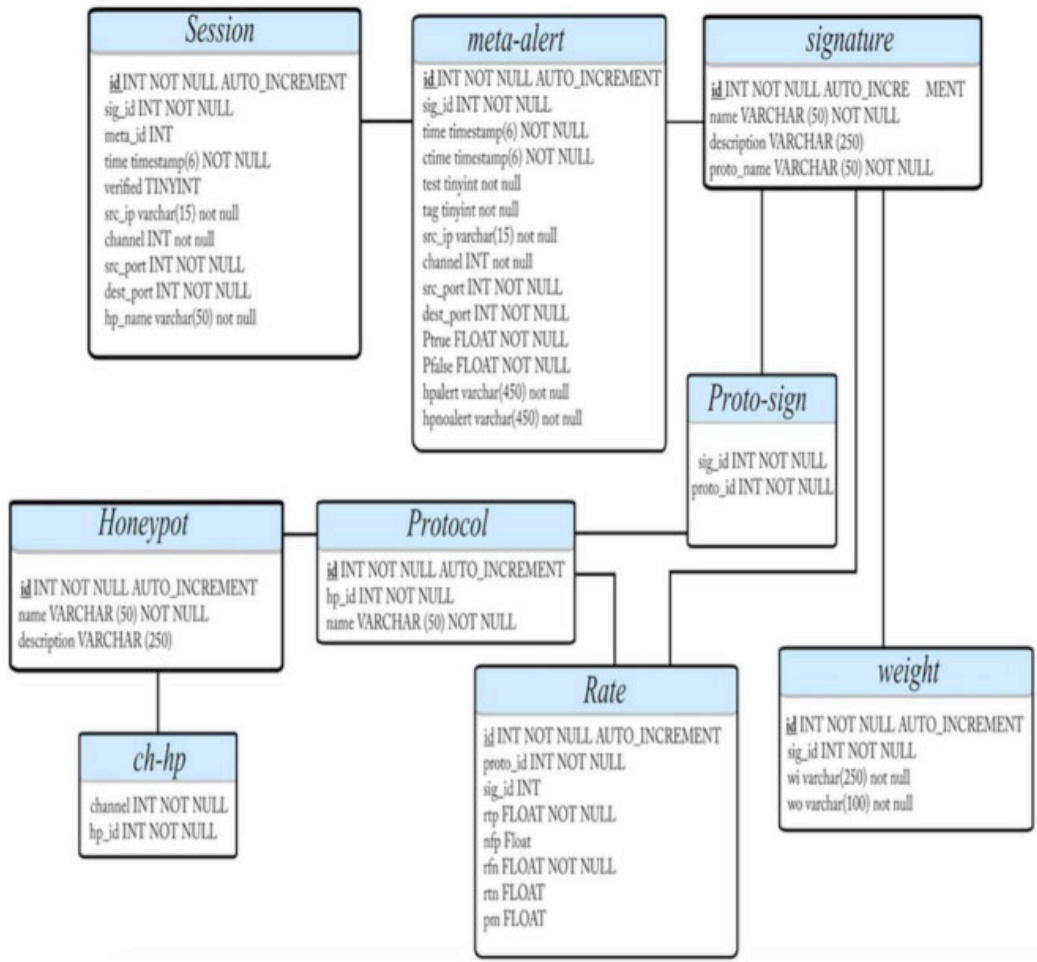

Figure 7. Class Diagram for the Framework database.

IV. Experimental Results

This section will be a test of our intelligent system; this test is based on a set of test data generated locally to test the functioning of both aggregation and verification components.

*A. Set of data used*

To test the operation of our probabilistic model and implemented multi-layer perceptron, we put a test scenario which assumes that we have a member site connected to the main server, and contains 3 sensors that can detect a specific signature attack on SSH protocol.

As we want to train our algorithm for a single attack signature and for 3 sensors Kippo, Suricata and Snort, we define the following parameters that describe the composition of the test traffic:

- The number of traces of attacks by signature: **15**.

- The number of normal traces by Protocol (equivalent to the number of normal traces by signature): **5**.

The table below describes testing dataset use.

*TABLE 2*
*DESCRIPTION OF TESTING DATASET USED*

|  | **Snort** | **Kippo** | **Suricata** |
|---|---|---|---|
| Number of testing malicious traces | 15 | 15 | 15 |
| Number of testing benevolent traces | 5 | 5 | 5 |
| Number of alerted sessions | 17 | 16 | 13 |
| Number of false alerts: False positive | 5 | 1 | 3 |
| Number of real alerts : True positive | 12 | 15 | 10 |
| Number of malicious traces but non-alerted : false negative | 3 | 0 | 5 |
| Number of benevolent traces and alerted : true negative | 0 | 4 | 2 |
| Rate of malicious alertsP(M=1) | 0.75 | 0.75 | 0.75 |

As illustrated in the table above, the data set consists of 46 alerts, with pre-verified attack sessions, in which the three sensors have committed false positives and false negatives.

By executing the aggregation algorithm, as result we had 20 meta-alerts summarizing alerts of attack sessions. The following figure shows the meta-alerts created and their associated fields:

Figure 8. Related meta-alerts associated to testing dataset

So we see that the aggregation component allowed us to reduce alerts to analyze, with a discount percentage of 56.5%.

### B. Testing framework capabilities:

Firstly, we launch the program to calculate different rates. The rates associated with each sensor are:

```
mysql> select honeypot.name,rtp,rfn,rtn,pm from rate,protocol,honeypot where rate.proto_id=protocol.id and protocol.hp_id=honeypot.id;
+---------+----------+-----------+-----+------+
| name    | rtp      | rfn       | rtn | pm   |
+---------+----------+-----------+-----+------+
| snort   | 0.823529 | 0.0666667 | 0.4 | 0.75 |
| kippo   |   0.9375 |         0 | 0.8 | 0.75 |
| suricata| 0.769231 |  0.333333 | 0.4 | 0.75 |
+---------+----------+-----------+-----+------+
3 rows in set (0,01 sec)
```

Figure 9. Calculated rates for each sensor

After calculating different rates, we move now to the training phase of the perceptron, in which we present just meta-alerts that are not complete, it means that meta-alerts where there was sensors that haven't generated alerts.

During this phase, we made several learning experiences, in each time we change both values: momentum and the rate of learning, to learn the values that will converge the multilayer perceptron as soon as possible to the optimal rates.

Firstly, we fixed the stop criteria of the training algorithm, and we gives to the performance index a maximum value of 0.02, and the maximum number of iterations is fixed at 20,000 iterations.

- **Exp1:** In this experiment we used a learning rate equal to '0.5' and a momentum equal to '0.1'.

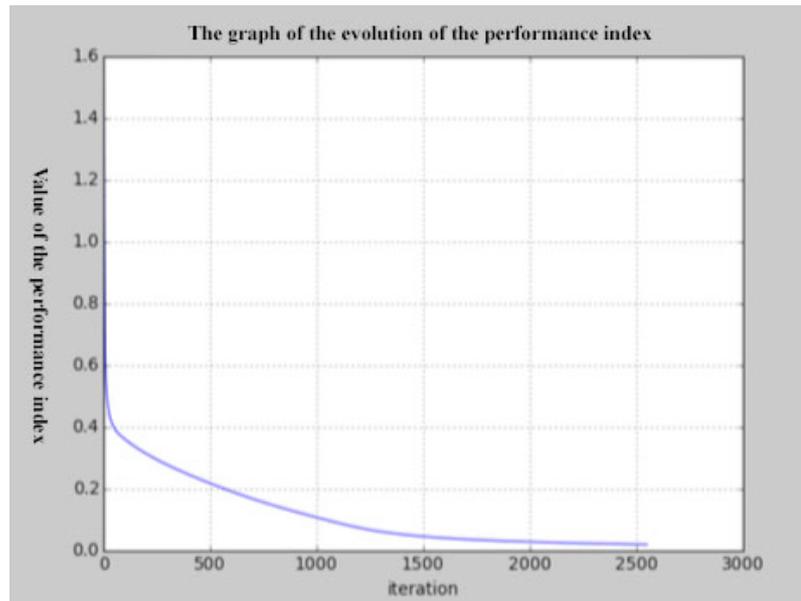

Figure 10. Graph evolution of the performance index in experience 1.

The performance index obtained for this experiment is equal to 0.0199 in 2551 iterations.

- **Exp 2**: In this experiment we used a learning rate equal to '0.8' and a momentum equal to '0.1'.

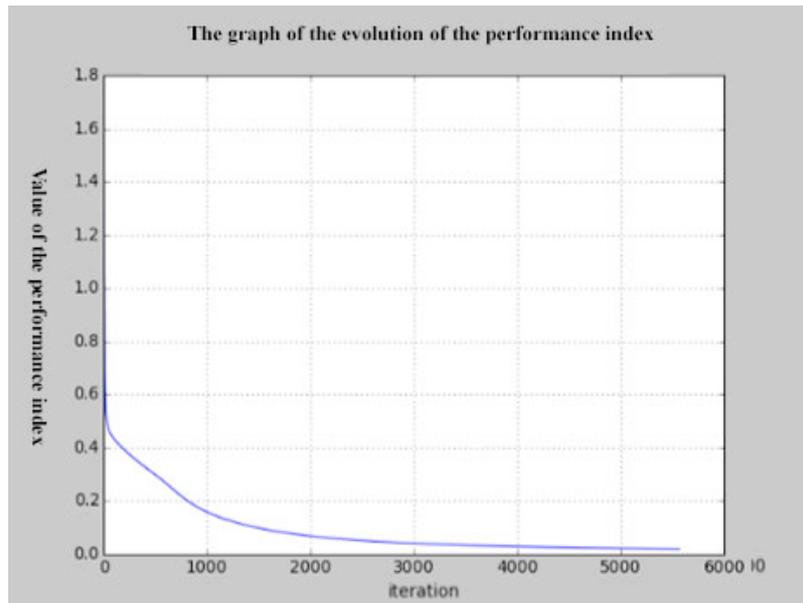

*Figure 11: Graph evolution of the performance index in experience 2.*

The performance index obtained for this experiment is equal to 0.0199 in 5574 iterations.

**Exp 3:** In this experiment we used a learning rate equal to '0.2' and a momentum equal to '0.1'.

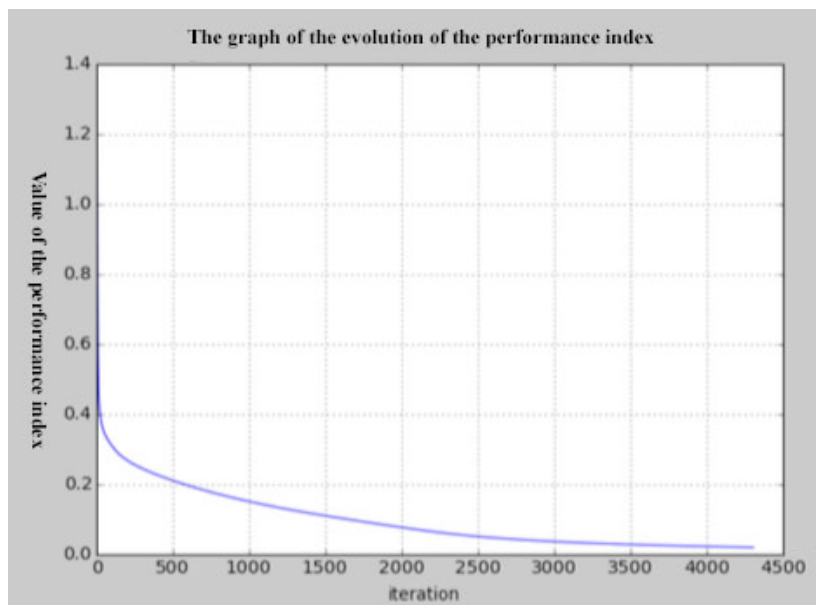

Figure 12. Graph evolution of the performance index in experience 3.

The performance index obtained for this experiment is equal to 0.0199 in 4307 iterations.

> **Analysis:**

In analyzing the results of the 3 previous experiences, we can see that with a learning rate equal to 0.5, we obtain a quicker learning. For that is fixed in the next experiences the learning rate to the value '0.5' and we changes the value of the momentum to try to minimize the performance index.

- **Exp 4:** In this experiment we used a learning rate equal to '0.5' and a momentum equal to '0.5'.

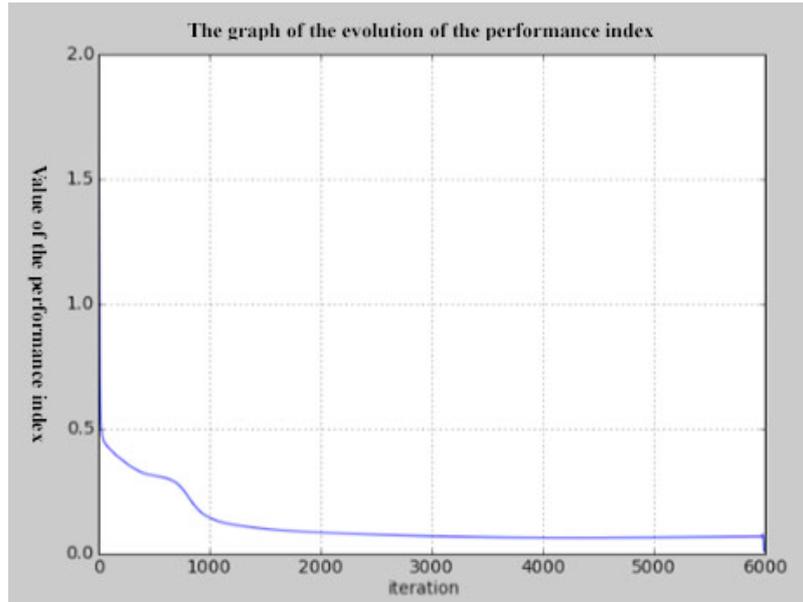

Figure 13. Graph evolution of the performance index in experience 4.

The performance index obtained for this experiment is equal to 0.0099 in 5993 iterations.

- **Exp 5:** In this experiment we used a learning rate equal to '0.5' and a momentum equal to '0.9'.

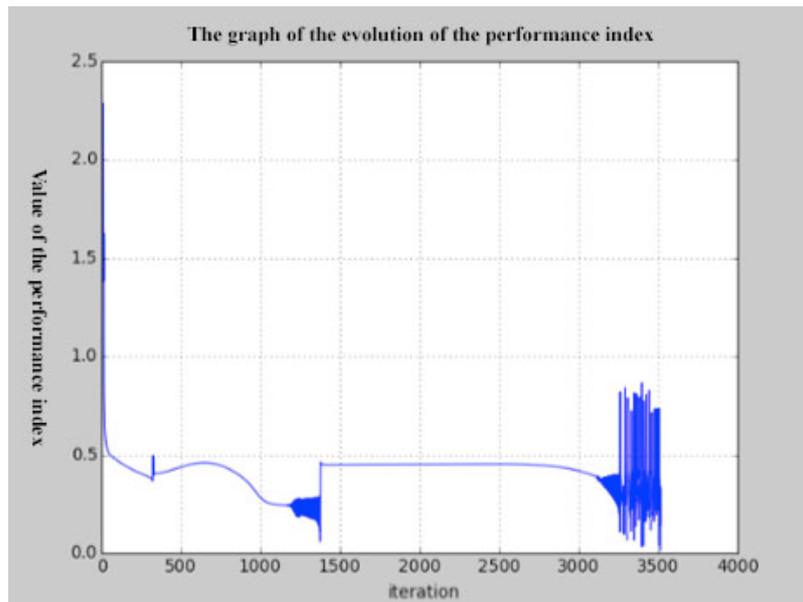

Figure 14. Graph evolution of the performance index in experience 5.

The performance index obtained for this experiment is equal to 0.0141 in 3517 iterations.

- **Exp 6:** In this experiment we used a learning rate equal to '0.5' and a momentum equal to '0.7'.

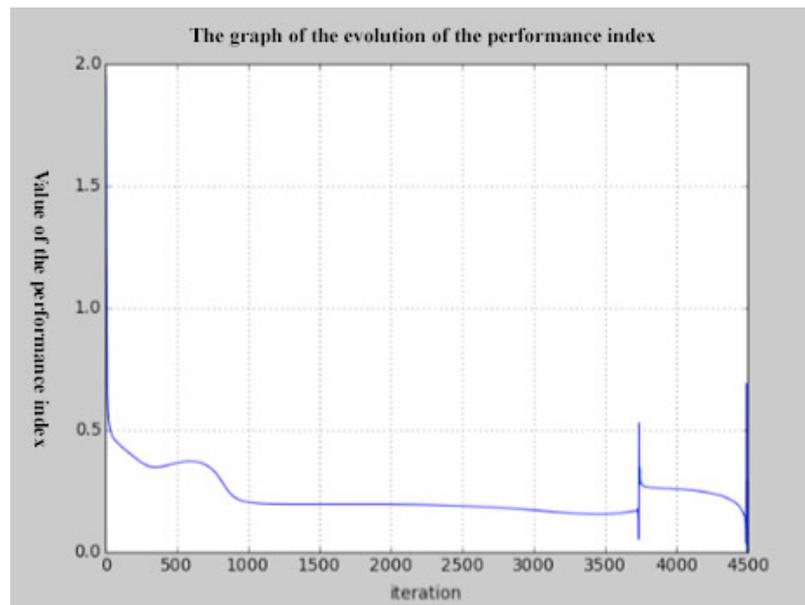

*Figure 15: Graph evolution of the performance index in experience 6.*

The performance index obtained for this experiment is equal to 0.0026 in 4495 iterations.

> **Analysis**

In the 3 recent experiments, we can see that the use of a momentum equal to '0.7' next to a learning rate equal to '0.5', gives a minimum value of the performance index of 0.0026, which implies that optimal rate calculated during the experiment 6 will be the best suited for a similar traffic to the training data.

V. Conclusion:

In this paper, we propose an automatic approach dedicated to reduce false positive and false negative from set of data collected by a distributed network of honeypots/IDS, and discover their characteristics. The analysis technique is based on machine learning method; Neural networks, our experimentation shows that we can reduce alerts to analyze, with a discount percentage of 56.5%.

The experimentation has also conducted to identify the minimum value of the performance index, which is the best suited for a similar traffic to the training data.

As a future work, we plan to expand the functionalities of our Framework, so that it can have a web application to better present the obtained results.

REFERENCES

[1]. Pouget F, Dacier M. 'Honeypot-based forensics'. Proceedings of Asia Pacific Information Technology Security Conference, Brisbane, Australia, 2004.

[2]. Alata E, Dacier M, Deswarte Y, et al. 'Collection and analysis of attack data based on honeypots deployed on the Internet.' Proceedings of First Workshop on Quality of Protection, Milan, Italy, 2005; 79–91.

[3]. Thonnard O, Dacier M. 'A framework for attack patterns' discovery in honeynet data'. Digital Investigation 2008; 8: S128–S139.

[4]. "About the honeynet project" [Online: Accessed 12 April 2016] http://old.honeynet.org/misc/project.html

[5]. ''Picviz : Data visualisation tool'' [Online: Accessed 25 Auguest 2016] https://www.honeynet.org/project/picviz

[6]. Seifert C, Komisarczuk P, Welch I. Identification of ma- licious Web pages with static heuristics. In Australasian Telecommunication Networks and Applications Confer- ence 2008, Adelaide.

[7]. ''Dionaea Project'' [Online: Accessed 25 Auguest 2016] https://github.com/rep/dionaea

[8]. "Kippo : an SSH Honeypot" [Online: Accessed 12 May 2016] https://github.com/desaster/kippo

[9]. ''p0f: Passive OS Fingerprinting tool'' [Online: Accessed 12 May 2016] http://lcamtuf.coredump.cx/p0f3/

[10]. Caswell, J. Beale, A. Baker, ''Snort IDS and IPS Toolkit'', Ed. SYNGRESS

[11]. ThreatStream, "Modern Honey Network," [Online: Accessed 12 April 2016] http://threatstream.github.io/mhn/

[12]. M. Schloesser, "Hpfeeds," https://github.com/rep/hpfeeds [Online: Accessed 12 April 2016]

[13]. J. V estergaard, "Mnemosyne," [Online: Accessed 12 April 2016] https://github.com/johnnykv/mnemosyne

[14]. MacQueen JB. Some methods for classification and analysis of multivariate observations. Proceedings of 5th Berkeley Symposium on Mathematical Statistics and Probability, Berkeley, University of California Press, 1967; 281–297.